\documentclass[lettersize,journal]{IEEEtran}
\usepackage{cite}
\usepackage{amsmath,amsfonts}
\usepackage{algorithmic}
\usepackage{array}
\usepackage[caption=false,font=footnotesize,labelfont=rm,textfont=rm]{subfig}
\usepackage{textcomp}
\usepackage{stfloats}
\usepackage{url}
\usepackage{verbatim}
\usepackage{graphicx}
\def\BibTeX{{\rm B\kern-.05em{\sc i\kern-.025em b}\kern-.08em
    T\kern-.1667em\lower.7ex\hbox{E}\kern-.125emX}}
\usepackage{balance}
\begin{document}
\title{Research on GEO SA-Bi SAR Imaging based on Joint Radar-Communications Waveform}
\author{Lian Meng,~\IEEEmembership{Member,~IEEE,} and Zhu Xu,~\IEEEmembership{Senior Member,~IEEE}
\thanks{(Corresponding author: Meng Lian). \par Lian Meng is with the School of Electronics and Information Engineering, Harbin Institute of Technology (Shenzhen), Shenzhen 518055, China (e-mail: lianmeng@hit.edu.cn). Zhu Xu is with the Department of Electrical Engineering and Electronics, The University of Liverpool, Liverpool L69 3BX, UK (e-mail: xuzhu@liverpool.ac.uk).}}

\markboth{Journal of \LaTeX\ Class Files,~Vol.~18, No.~9, September~2020}%
{How to Use the IEEEtran \LaTeX \ Templates}

\maketitle

\begin{abstract}
Joint radar-communications (JRC) technology has attracted massive attention for decades, since it can effectively utilize allocated spectral resources by sharing frequency bands in increasingly crowded environments. In addition, the growing demand for hardware platform sharing which benefits both functionalities motivates more cooperation between radar and communication systems. In order to achieve the coexistence of sensing and communicating operations, joint systems should be designed to perform both tasks simultaneously. Developing a joint radar-communications waveform which is suitable for both functions is extremely crucial for this type of co-design, as it not only decreases spectral impact, but also benefits performances of both systems mutually. In this paper, a joint radar-communications waveform is utilized to perform GEO SA-Bi SAR imaging and wireless communication simultaneously. We also design a joint radar-communications receiver in this context to demonstrate feasibility of achieving both sensing and signaling with GEO SA-Bi SAR system.
\end{abstract}

\begin{IEEEkeywords}
Joint Radar-Communications Waveform, Geosynchronous Spaceborne/Airborne Bistatic Synthetic Aperture Radar, Ship Target Imaging.
\end{IEEEkeywords}

\section{Introduction}
\IEEEPARstart{D}{ue} to the benefits of wide imaging area and distributed configuration, geosynchronous spaceborne/airborne bistatic synthetic aperture radar (GEO SA-Bi SAR) system have attracted massive attentions from a diverse range of remote sensing applications in recent years \cite{ZCui1,ZCui2,DLi}. However, wireless communication systems now compete with radar systems for bandwidth in frequency domain owing to the increasing demand for communication devices. Consequently, spectrum is becoming more and more congested and the performance of GEO SA-Bi SAR will be limited by the increasingly complexity of electromagnetic environment. A emerging strategy for solving this problem is joint radar-communication technology which allows for a integration of both systems and mutually reinforcing functionalities \cite{AHer}. Therefore, a number of researchers focus on the study of JRC and plenty of articles have been published in recent years \cite{ZFeng}.

Despite, there is strong hardware similarity between SAR and communication systems, the principles of them are quite different \cite{BPaul}. In SAR system, the original transmit signal is known for receiver which utilizes it to estimate target characteristics by matched filtering. However, the signal transmitted by communication system is usually random because of the modulated information, and the channel state information can be obtained previously through data carried in channel estimation \cite{CAJackson}. Accordingly, SAR receiver needs precise form of transmitted signal as reference to achieve focused images (which can be regarded as a sort of channel information), but communication receiver only needs primary information of channel, which may be partly omitted sometimes. As a result, problems, such as electromagnetic interference suppression and functional reconfiguration, need to be settled in JRC design \cite{JWang}. 

In order to achieving JRC, joint waveform and transceiver design must be concerned, which are great challenges. As mentioned above, the main reason is different requirements of radar and communication are induced by their own operation principles. According to the shared spectrum access for radar and communications programs \cite{GMJacyna,ARChiriyath,CDRichmond}, these requirements may lead to conflicts in the process of  waveform design. Particularly, SAR needs higher signal-to-clutter ratio (SCR) and signal that with larger time-bandwidth product to ensure the performance of imaging, which makes the contradiction more intense. Some researchers addressed this problem by using linear frequency modulated (LFM) signal and spread spectrum (SS) signal. In \cite{SJXu1}, Direct sequence spread spectrum (DSSS) technique was introduced to ensure the orthogonality between sensing and communication signals, which increases the security and robustness of JRC system. Furthermore, a combination of DS and ultra wide band (UWB) radars was proposed in \cite{SJXu2} to spread spectrum of both sensing and communication signals by adopting different pseudo-noise code respectively, which suppresses jamming among different functions. Based on LFM, minimum shift keying (MSK) and continuous phase modulation (CPM), some novel compatible waveforms were designed through modifying mapping codebook to decrease the loss induced by attenuation of the power amplifier and suppress the interference brought by other out-of-band devises \cite{XBChen,YZhang1}. However, the parameter adjustment of these approaches is not flexible, and the data load and transmission rate are low. Moreover, accurate signal separation is needed or the orthogonality of integrated signal must be guaranteed, which lead to high costs and complexity. As orthogonal frequency division multiplexing (OFDM) waveform which enables low sidelobe, high Doppler tolerance and information transmission capacity, is widely applied in communication, it has attracted great attentions in the development of JRC. The design of a JRC system which utilizes OFDM waveform to achieve both radar sensing and communication functions simultaneously, was first proposed in \cite{CSturm1}. Based on the study in \cite{CSturm2}, OFDM based method were proved to have a very flexible detection range in radar imaging tasks. Moreover, an approach that employs standard OFDM signal to estimate the velocity of targets was built in \cite{YLSit}. However, the extremely high peak-to-average power ratio (PARPA) of standard OFDM communication signal will leads to inevitable distortion in transmitted waveform and incomplete usage of radar’s power amplifier, which reduces the detection range of radar sensing considerably \cite{YZhang2}. Besides, the above methods to realize JRC mainly focus on the signal level integration to achieve both radar sensing and communication functions, while the design of joint receiver which shrinks the volume of JRC system is not considered sufficiently.

The main contributions of this letter are three-folds:
\begin{list}{}{}
\item{1) This is the first research on JRC design for GEO SA-Bi SAR system and it enables both ship target imaging and communication functions without increasing the cost and volume of original GEO SA-Bi SAR system significantly.}
\item{2) Comparing with existing JRC schemes, the approach in this paper employs one single JRC waveform and a JRC receiver which are designed to achieve both GEO SA-Bi SAR imaging and communication goals simultaneously. It avoids mutual interference induced by insufficient signal separation or non-orthogonality of integrated signal, which improves the performance of entire JRC system.}
\item{3) Comparing with the SAR imaging methods which mainly focus on static scene in JRC researches, the GEO SA-Bi SAR imaging approach in this letter concerns ship targets that influenced by ocean waves, and contaminated by background clutter under rough sea conditions. It significantly improves the quality of nonstationary targets imaging in the context.}
\end{list}

Experiments on the simulated data are performed to demonstrate the correctness and effectiveness of the proposed JRC approach. Comparing with other JRC methods, the image resolution and transmission rate are both improved significantly.  

\section{Signal Model}
\begin{figure}[!t]
\centering
\includegraphics[width=2.5in]{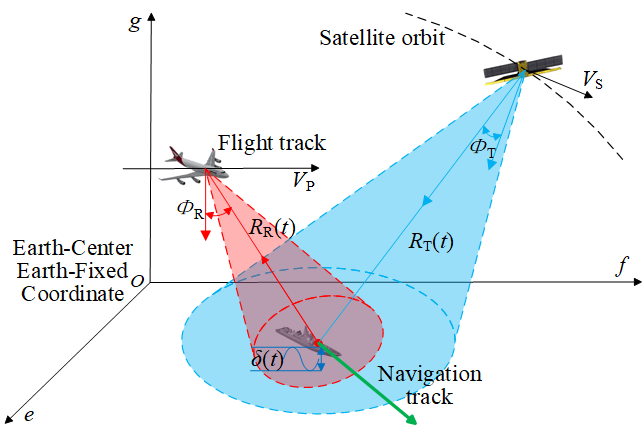}
\caption{Problem geometry for GEO SA-Bi SAR based JRC system. GEO transmitter focuses its antenna on a ship target within a ocean scene. Elevation and azimuth angles are defined with respect to the ship target. The velocity and acceleration of ship target are divided into radial and along-track components with respect to each platform, respectively. The ship target also moves vertically due to a perturbation induced by ocean wave.}
\label{f1}
\end{figure}

\noindent The problem geometry is indicated in Fig. \ref{f1}, which shows the geometrical architecture of a GEO SA-Bi SAR system which consists of a GEO SAR utilized to emit JRC signals and an airborne SAR equipped with a JRC receiver. The GEO SAR transmits JRC signals towards an ocean scene where a ship target is cruising. The GEO SAR is designated as the transmitter, while the airborne SAR as the receiver. The transmitter transmits JRC signals through the target contained scene to the receiver continuously. This process is repeated for different transmit-receive configurations within the whole synthetic aperture time. The final reception consists of echo signals for each configuration, and is decoded to achieve messages from transmitter. Based on the obtained messages, the precise form of transmitted signal are then derived and utilized for SAR imaging. 

Generally, the transmitter mounted on a GEO satellite moves along a predetermined orbit with the velocity \(V_{\text{S}}\). Under the assumption of flat earth surface, the effective velocity of the satellite is approximately equal to \(V_{\text{eff\_S}} \approx \sqrt{V_{\text{S}}V_{\text{gr\_S}}}\), and the absolute and ground velocities of the satellite in Earth center Earth fixed coordinate system (ECEF) are presented by \(V_{\text{S}}\) and \(V_{\text{gr\_S}}\) , respectively \cite{EMakhoul}. According to \cite{YJiang}, the range history of a marine moving target respect to the satellite can be written as: 
\begin{equation}
\label{eq1}
\begin{split}
R_{\text{T}}(t) & \approx R_{\text{T0}}+\delta(t)+(V_{\text{S}}\Phi_{\text{T}}+V_{\text{rad\_T}})t \\
& +0.5(V_{\text{eff\_S}}^{2}/R_{\text{T0}}-2V_{\text{S}}V_{\text{al\_T}}/R_{\text{T0}}+A_{\text{rad\_T}})t^{2}.
\end{split}
\end{equation}
In the same way, the range history of this marine moving target respect to the airplane can be written as:
\begin{equation}
\label{eq2}
\begin{split}
R_{\text{R}}(t) & \approx R_{\text{R0}}+\delta(t)+(V_{\text{P}}\Phi_{\text{R}}+V_{\text{rad\_R}})t \\
& +0.5(V_{\text{eff\_P}}^{2}/R_{\text{R0}}-2V_{\text{P}}V_{\text{al\_R}}/R_{\text{R0}}+A_{\text{rad\_R}})t^{2}.
\end{split}
\end{equation}
where \(t\) denotes slow time. \(R_{\text{T0}}\) and \(R_{\text{R0}}\) are the initial distances of the marine moving target respect to the satellite and airplane respectively. \(\Phi_{\text{T}}\) and \(\Phi_{\text{R}}\) represent the squint angle of the GEO SAR and the airborne SAR. The absolute and effective velocities of the airplane in ECEF coordinate system are presented by \(V_{\text{P}}\) and \(V_{\text{eff\_P}}\), and \(V_{\text{eff\_P}}=V_{\text{P}}=V_{\text{gr\_P}}\) (\(V_{\text{gr\_P}}\)  is the ground velocity of airplane). \(V_{\text{rad\_T}}\)(\(V_{\text{rad\_R}}\)) and \(A_{\text{rad\_T}}\)(\(A_{\text{rad\_R}}\)) represent the radial velocity and radial acceleration of target along \(R_{\text{T}}\)(\(R_{\text{R}}\)), respectively. \(V_{\text{al\_T}}\)(\(V_{\text{al\_R}}\)) denotes the along-track velocity of target parallel to \(V_{\text{S}}\)(\(V_{\text{P}}\)). The term \(\delta(t)\) appears on the right side of both above equations indicates a perturbation induced by ship heave motion, which is assumed to be a superposition of simple harmonic terms and can be expressed as:
\begin{equation}
\label{eq3}
\delta(t) \approx \sum_{k=1}^{n}(p_{k}\text{cos}(\mit\Omega_{k}t)+q_{k}\text{sin}(\mit\Omega_{k}t)).
\end{equation}
where \(\mit\Omega_{k}\) is the counter wave frequency, while \(p_{k}\) and \(q_{k}\) are the quadrature components of amplitude. Therefore, the bistatic range history of a marine moving target in GEO SA-Bi SAR  framework can be expressed as:
\begin{equation}
\label{eq4}
R_{\text{B}}(t)=R_{\text{T}}(t)+R_{\text{R}}(t).
\end{equation}

A standard Quadrature phase shift keying (QPSK) communication waveform is assumed to be used to illuminate the scene in this context. Let \(b1_{i}\) and \(b2_{i}\) represent the even and odd sequences of the original data stream, respectively. The base band pulse is considered as rectangular pulse, so \(g(t)=g_{\text{R}}(t,T_{\text{S}})\). Therefore, the two rectangular pulse sequences which carry the even and odd data flow respectively, can be defined as:
\begin{equation}
\label{eq5}
\begin{split}
p1_{\text{D}}(t) & =\sum_{i=-\infty}^{\infty}b1_{i}g(t-iT_{\text{S}})=b1_{i}*g(t)\\
p2_{\text{D}}(t) & =\sum_{i=-\infty}^{\infty}b2_{i}g(t-iT_{\text{S}})=b2_{i}*g(t)
\end{split}
\end{equation}
where \(T_{\text{S}}\) is the symbol interval of original data stream. When QPSK is utilized in amplitude modulation, the bandpass signal can be presented as:
\begin{equation}
\label{eq6}
s_{\text{BP}}(t)=\sqrt{E_{\text{B}}/T_{\text{B}}}[p1_{\text{D}}(t)\text{cos}(2\pi f_{\text{c}}t)-p2_{\text{D}}(t)\text{sin}(2\pi f_{\text{c}}t)].
\end{equation} 
where \(T_{\text{B}}\) is the symbol interval in \(b1_{i}\) and \(b2_{i}\), so \(T_{\text{S}}=2T_{\text{B}}\). \(E_{\text{B}}\) represents the amount of enegy costed by transmitting one bit information, which satisfies \(\int_{0}^{T_{\text{S}}}s_{\text{BP}}(t)^{2}\text{d}t=2E_{\text{B}}\). The baseband signal can be written as:
\begin{equation}
\label{eq7}
s_{\text{LP}}(t)=\sqrt{E_{\text{B}}/T_{\text{B}}}[p1_{\text{D}}(t)+jp2_{\text{D}}(t)].
\end{equation}

After demodulation, the signal captured along the range cell migration (RCM) line in the slow time-slant range domain can be written as:
\begin{equation}
\label{eq8}
\mathbf{y}(t)=a(t)s_{\text{LP}}(t-\tau)\text{exp}[-j2\pi R_{\text{B}}(t)/\lambda]+\mathbf{e}(t).
\end{equation}
where \(\tau\) is the time delay experienced by reflected signal, and \(\tau=R_{\text{B}}/c\) which can be calculated by Eq. (\ref{eq1})-(\ref{eq4}), \(c\) is the speed of light, \(\lambda\) is the carrier wavelength, the additional term \(\mathbf{e}(t)\) denotes sea clutter and noise. Generally, the unknown amplitude term \(a(t)\) depends on the type of target, and the phase is assumed to depend only on the corresponding bistatic range history. 

Since the range resolution of SAR mainly depends on the bandwith of transmitted signal, large time-bandwidth product signal, such as LFM signal, is commonly used in SAR system. However, the data load and transmission rate of LFM signal are low as mentioned previously. Therefore, we adopt direct sequence spread spectrum (DSSS) technique combined with QPSK to overcome these problems. According to Eq. (\ref{eq6}), the bandwidth of tansmitted signal is extremely narrow,which would lead to inevitable low resolution in range direction. To increase the bandwidth, Kasami spread sequence is appled to sub data stream \(b1_{i}\) and \(b2_{i}\) before they flow into raised cosine rolling-off shaping filter, respectively. This process can be expressed as:
\begin{equation}
\label{eq9}
\begin{split}
p1_{\text{S}}(t_{s}) &= p1_{\text{D}}(t_{s}) \oplus kasa(t_{s})\\
p2_{\text{S}}(t_{s}) &= p2_{\text{D}}(t_{s}) \oplus kasa(t_{s})
\end{split}
\end{equation}
where \(p1_{\text{S}}(t_{s})\) and \(p2_{\text{S}}(t_{s})\) indicate the sequences after DSSS, \(kasa(t_{s})\) is the Kasami spread sequence, \(t_{s}\) is the spreaded form of \(t\) which is decided by oversampling rate and length of Kasami sequence, \( \oplus \) represents XOR operation. Besides, DSSS can also effectively suppress inter-symbol interference (ISI) in wireless communication. In order to obtain the original sequence, the demodulated signal is despreaded with the same Kasami sequence in JRC receiver which will be introduced in the next section.   

\section{Joint Reveiver Design}
\noindent A JRC receiver that simultaneously performs communication and GEO SA-Bi SAR ship target imaging is developed and employed in the context of this research. This JRC receiver first extracts and decodes messages that transmitted by GEO SAR. Then, the decoded messages are utilized to rebuild the  waveforms of transmitted signals. Finally, these estimations of the original transmitted signals are used as reference signals to perform GEO SA-Bi SAR imaging.

Ignoring the Doppler shifts induced by different sub-channels, the received signal Eq. (\ref{eq8}) can be simplified as:
\begin{equation}
\label{eq10}
\mathbf{y}(t)=\mathbf{h}(t)*s_{\text{LP}}(t)+\mathbf{e}(t).
\end{equation}
where \(\mathbf{h}(t)\) is the simplified channel model. It is shown that Eq. (\ref{eq10}) is a standard communication model and can be regarded as a rational approximation without Doppler spread of channel. According to \cite{AHerschfelt}, an equalizer which utilizes Toeplitz completion to estimate the autocovariance matrix of the received signal \(\mathbf{y}(t)\) can be established with least-squares Wiener filter. Therefore, the transmission signal \(\hat{s}_{\text{LP}}(t)\) can be estimated by applying this equalizer. Consequently, the estimated channel \(\hat{\mathbf{h}}(t)\) can be achieved with \(\hat{s}_{\text{LP}}(t)\). After achieving the estimated transmissions, the waveforms of transmitted signals are then developed and utilized as references for GEO SA-Bi SAR imaging. 

There are two main challenges in GEO SA-Bi SAR ship target imaging with JRC waveforms. One is that the reference signal used in range compression is usually random because of the loaded data. As a result, the performance of range compression will extremely decrease due to deficient matched filtering. However, \(\hat{s}_{\text{LP}}(t)\) can be achieved through applying the above process, which provides a solution to this problem. It means that the reference signal \(\hat{s}_{\text{BP}}(t)\) which is used as matching function in range compression can be estimated easily. Another challenge is that the motion of a ship target includes heave components induced by ocean waves, which will introduce periodic components in Doppler history and lead to azimuth defocus in SAR images. After range compression, the signal captured along RCM line in slow time-slant range domain can be written as:
\begin{equation}
\label{eq11}
\mathbf{x}(t)=a(t)\text{exp}[-j2\pi R_{\text{B}}(t)/\lambda]+\mathbf{e}(t).
\end{equation}
According to Eq. (\ref{eq1}), Eq. (\ref{eq2}) and Eq. (\ref{eq11}), and removing the azimuth Doppler rate of stationary target for simplifying analysis, the signal model \(\mathbf{x}(t)\) can be expanded as:
\begin{equation}
\label{eq12}
\begin{split}
\mathbf{x}(t) &=a(t)\text{exp}[-j2\pi(R_{\text{T0}}+R_{\text{R0}}+2\delta(t))/\lambda]\\
&\cdot\text{exp}[-j2\pi(V_{\text{S}}\Phi_{\text{T}}+V_{\text{S}}\mu_{\text{T}}+V_{\text{P}}\Phi_{\text{R}}+V_{\text{P}}\mu_{\text{R}})t/\lambda]\\
&\cdot\text{exp}[-j\pi(V_{\text{S}}^{2}(-2\nu_{\text{T}}+\eta_{\text{T}})/R_{\text{T0}})t^{2}/\lambda]\\
&\cdot\text{exp}[-j\pi(V_{\text{P}}^{2}(-2\nu_{\text{R}}+\eta_{\text{R}})/R_{\text{R0}})t^{2}/\lambda]
\end{split}
\end{equation}
where \(\mu_{\text{T}}=V_{\text{rad\_T}}/V_{\text{S}}\), \(\mu_{\text{R}}=V_{\text{rad\_R}}/V_{\text{P}}\), \(\nu_{\text{T}}=V_{\text{al\_T}}/V_{\text{S}}\), \(\nu_{\text{R}}=V_{\text{al\_R}}/V_{\text{P}}\), \(\eta_{\text{T}}=R_{\text{T0}}A_{\text{rad\_T}}/V_{\text{S}}^{2}\), \(\eta_{\text{R}}=R_{\text{R0}}A_{\text{rad\_R}}/V_{\text{P}}^{2}\) are normalized motion parameters. From Eq. (\ref{eq12}), the Doppler history can be derived as:
\begin{equation}
\label{eq13}
\begin{split}
f_{\text{D}}(t) &\approx -[(V_{\text{S}}\Phi_{\text{T}}+V_{\text{rad\_T}}+V_{\text{P}}\Phi_{\text{R}}+V_{\text{rad\_R}})\\
&+(V_{\text{S}}^{2}(-2\nu_{\text{T}}+\eta_{\text{T}})/R_{\text{T0}}+V_{\text{P}}^{2}(-2\nu_{\text{R}}+\eta_{\text{R}})/R_{\text{R0}})t\\
&+2\sum_{k=1}^{n}(q_{k}\mit\Omega_{k}\text{cos}(\mit\Omega_{k}t)-p_{k}\mit\Omega_{k}\text{sin}(\mit\Omega_{k}t))]
\end{split}
\end{equation}
From Eq. (\ref{eq13}), it can be seen that the Doppler history consists of the linear components caused by the horizontal motion of ship target and the periodic components caused by the heave motion due to ocean waves. 

\begin{figure}[!t]
\centering
\includegraphics[width=2.5in]{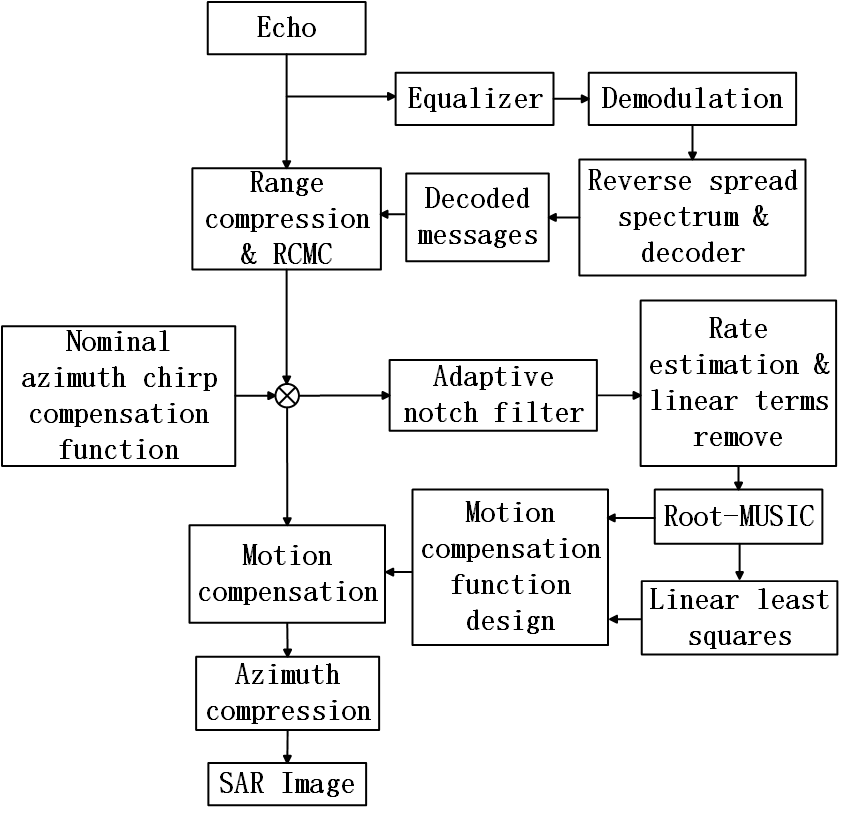}
\caption{The flow chart of the proposed JRC receiver strategy.}
\label{f2}
\end{figure}

In our previous work \cite{YCJiang}, we developed an adaptive approach which combines adaptive notch filtering (ANF) with root multiple signal classification (Root-MUSIC) algorithm to address this problem. Firstly, we use ANF to extract the Doppler history of a ship target, which is a composite signal consists of linear and periodic components. Then, the linear terms can be removed after estimating the slop of instantaneous Doppler frequency. Consequently, the residual Doppler frequency is the period terms induced by ocean waves. Therefore, the frequencies of the waves can be estimated by the Root-MUSIC algorithm. After estimating the wave frequencies, the amplitudes of the waves can be estimated by linear least squares method. At last, the motion caused by propulsion and waves can be compensated by applying the designed functions based on the estimated parameters. The strategy of the proposed JRC receiver is described in Fig. \ref{f2}.

\section{Simulation Results}
\noindent In this section, the received JRC signal in terms of GEO SA-Bi SAR moving ship target echo influenced by sea clutter is simulated. Both a point target and a dot matrix ship are imaged to demonstrate the feasibility of the JRC receiver. The simulation parameters are detailed in Table \ref{tab1}, \ref{tab2} and \ref{tab3}.

\begin{table}
\begin{center}
\caption{System Parameters}
\label{tab1}
\begin{tabular}{| l | l |}
\hline
System Parameters & Value\\
\hline
Bandwidth & 100~MHz\\
\hline
Carrier Frequency & 10~GHz\\ 
\hline
Modulation & QPSK\\
\hline 
Spread Sequence & Kasami\\
\hline
Encoder & Turbo\\
\hline
Code Rate & 0.125\\
\hline
PRF & 863~kHz\\
\hline
Synthetic Aperture Length & $\sim$273.1~m\\
\hline
\end{tabular}
\end{center}
\end{table}

\begin{table}
\begin{center}
\caption{Platform Parameters}
\label{tab2}
\begin{tabular}{| l | l | l |}
\hline
Platform Parameters & Transmitter & Receiver\\
\hline
Altitude & 36,000~km & 8~km\\
\hline
Speed & 2,300~m/s & 200~m/s\\ 
\hline
Elevation & 0.6~deg & 25.7~deg\\
\hline 
Bearing & 0~deg & 4~deg\\
\hline
\end{tabular}
\end{center}
\end{table}

\begin{table}[!t]
\begin{center}
\caption{Target Parameters}
\label{tab3}
\begin{tabular}{| l | l |}
\hline
Target Parameters & Value\\
\hline
Altitude & 0~m\\
\hline
Speed & 10.5~m/s\\ 
\hline
Acceleration & 1.0~m/\(\text{s}^{2}\)\\
\hline 
Amplitude of Reflection Coefficients  & 1.1\\
\hline
Direction Respect to Airplane's Velocity & 30.2~deg\\
\hline
Amplitude of \(\delta(t)\) & 1.6~m\\
\hline
Period of \(\delta(t)\) & 3.5~s\\
\hline
\end{tabular}
\end{center}
\end{table}

\subsection{Point Target}
\begin{figure}[!t]
\centering
\subfloat[Range response]{\includegraphics[width=2.5in]{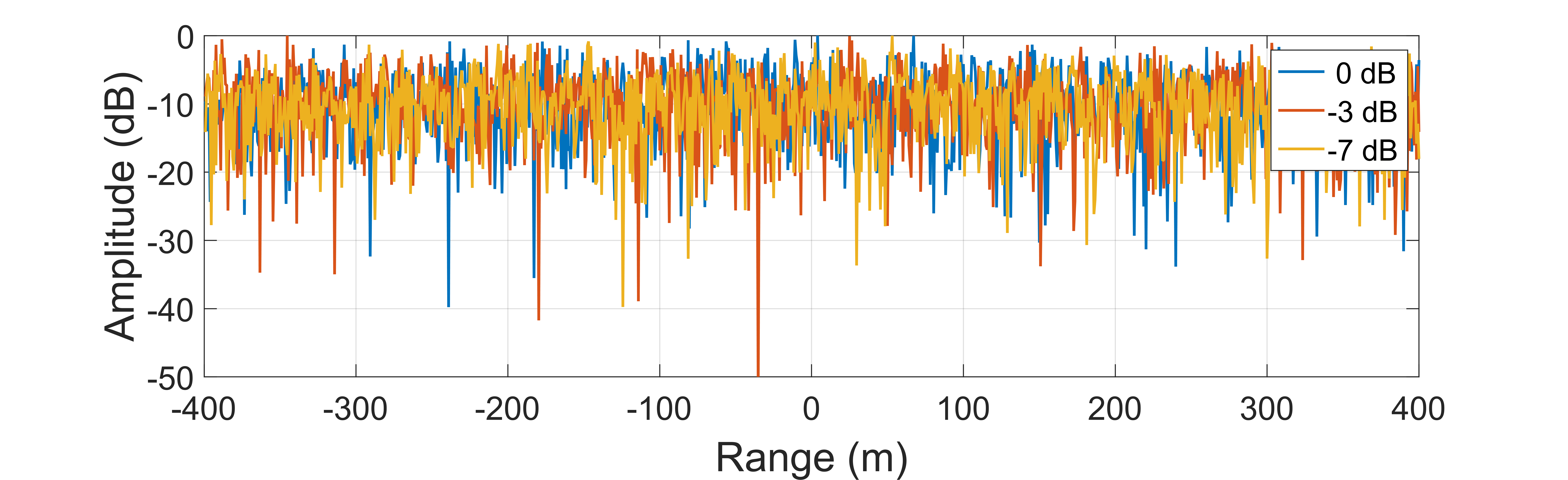}%
\label{fig3_a}}
\vfil
\subfloat[Azimuth response]{\includegraphics[width=2.5in]{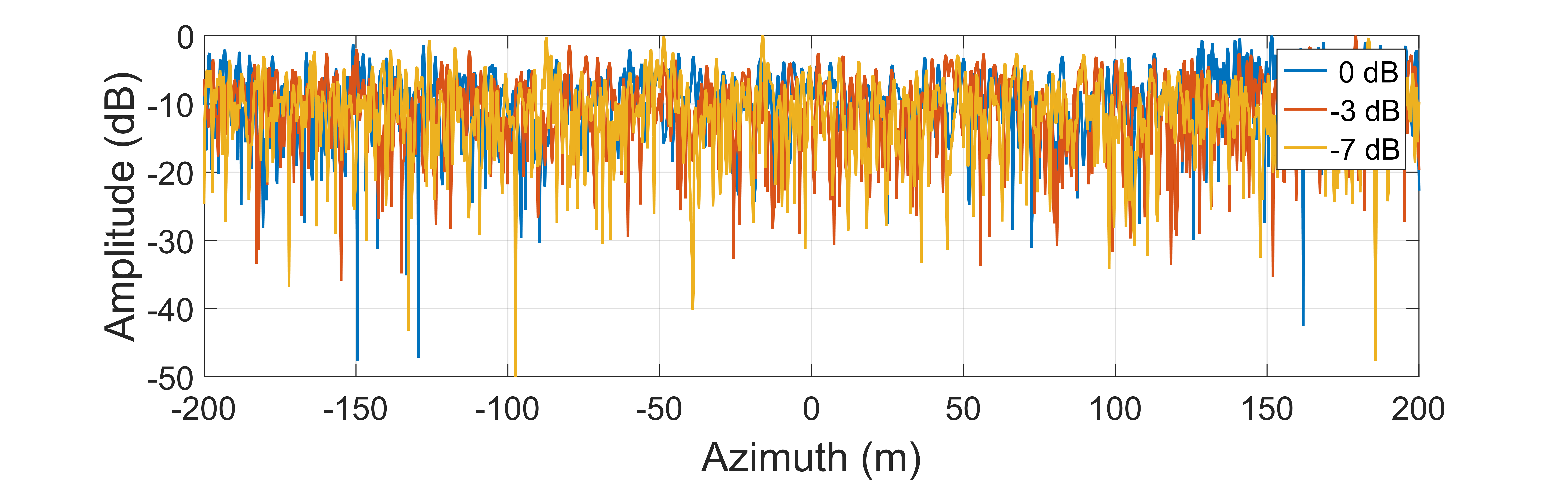}%
\label{fig3_b}}
\caption{Simulation result of point target imaging without motion compensation in different SNR. (a) Range response of the point target located in scene center. (b) Azimuth response of the point target located in scene center.}
\label{f3}
\end{figure}

\begin{figure}[!t]
\centering
\subfloat[Range response]{\includegraphics[width=2.5in]{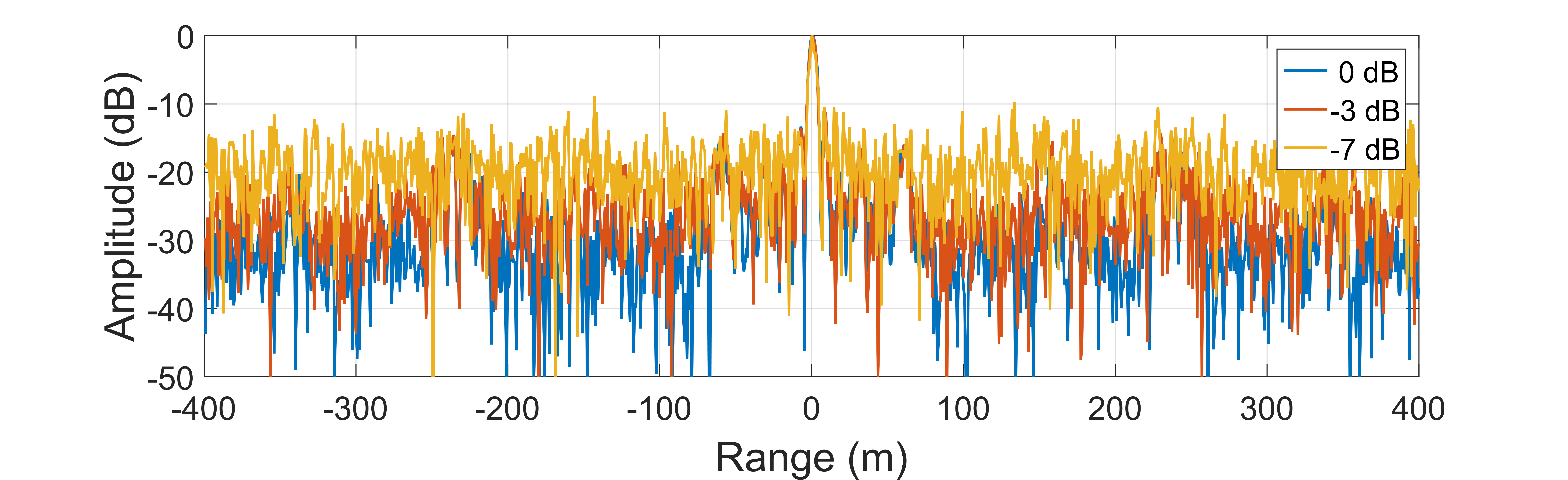}%
\label{fig4_a}}
\vfil
\subfloat[Azimuth response]{\includegraphics[width=2.5in]{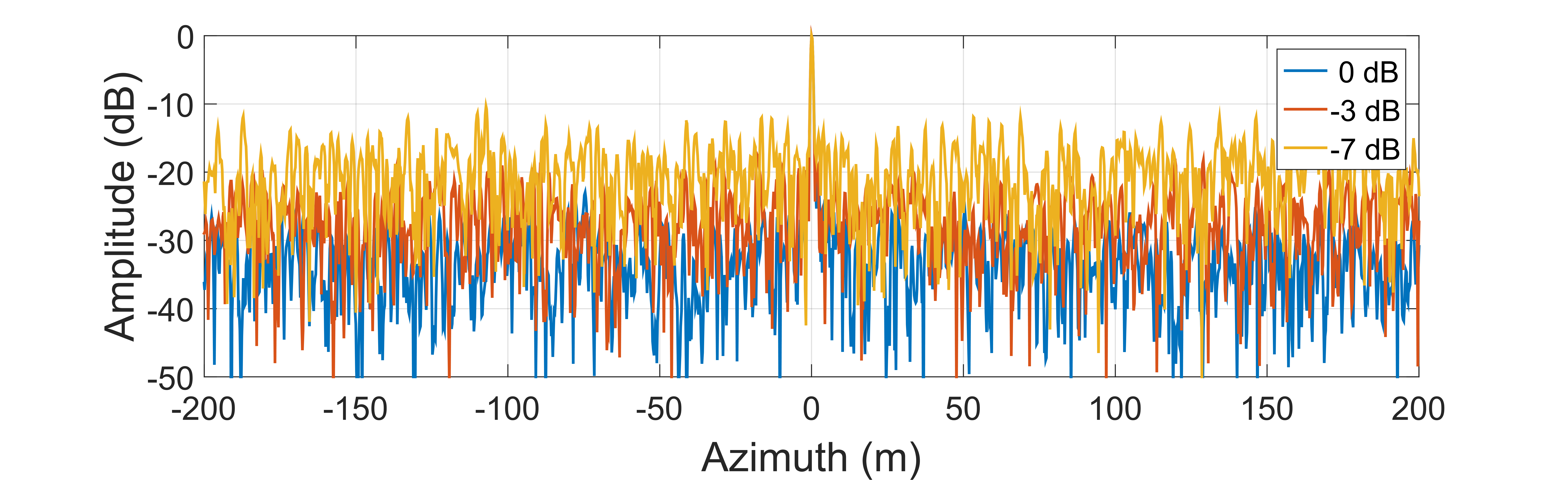}%
\label{fig4_b}}
\caption{Simulation result of point target imaging with motion compensation in different SNR. (a) Range response of the point target located in scene center. (b) Azimuth response of the point target located in scene center.}
\label{f4}
\end{figure}

To evaluate the performance of the proposed JRC strategy, the imaging results of a point target located in the center of imaging field are shown in Fig. \ref{f3} and Fig. \ref{f4}. The whole JRC system which consists of one GEO SAR transmitter and one airborne SAR receiver utilizes a QPSK communication waveform with a Kasami spread sequence to sense the nonstationary point target. The point target simulation result allows for a better system parameters calibration and setting with a given waveform. Besides, the different behaviors and characteristics of GEO SA-Bi SAR imaging strategies can be quantified.


\subsection{Ship Target}
\begin{figure}[!t]
\centering
\subfloat[Dot matrix model]{\includegraphics[width=2.5in]{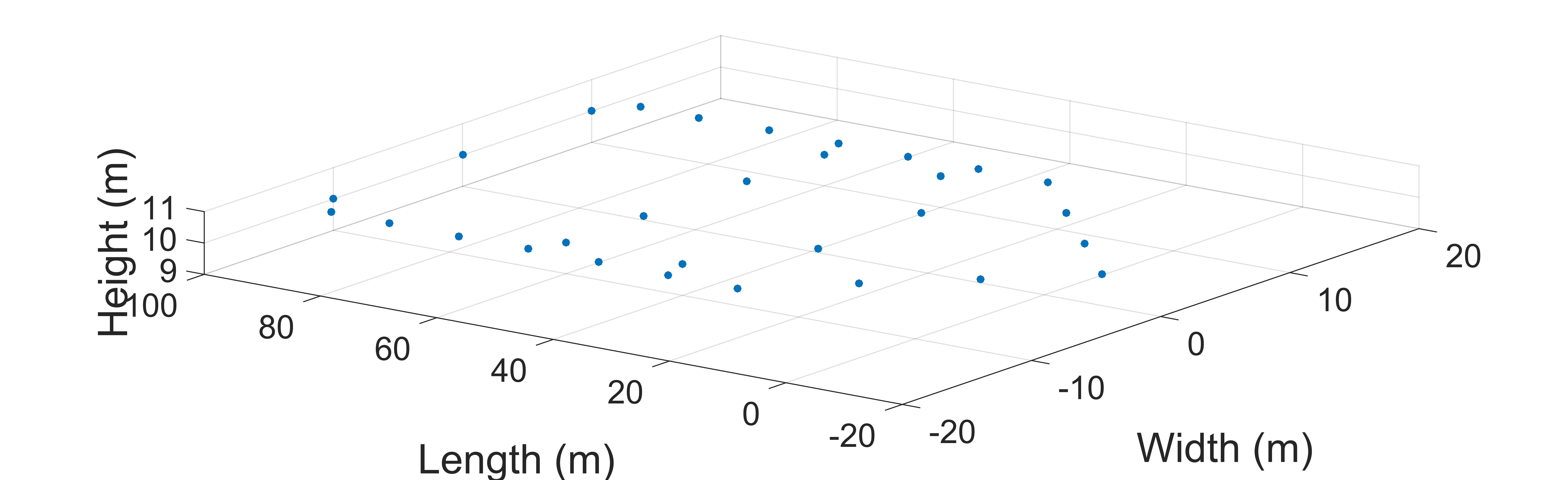}%
\label{fig5_a}}
\vfil
\subfloat[Result by non-compensation method]{\includegraphics[width=2.5in]{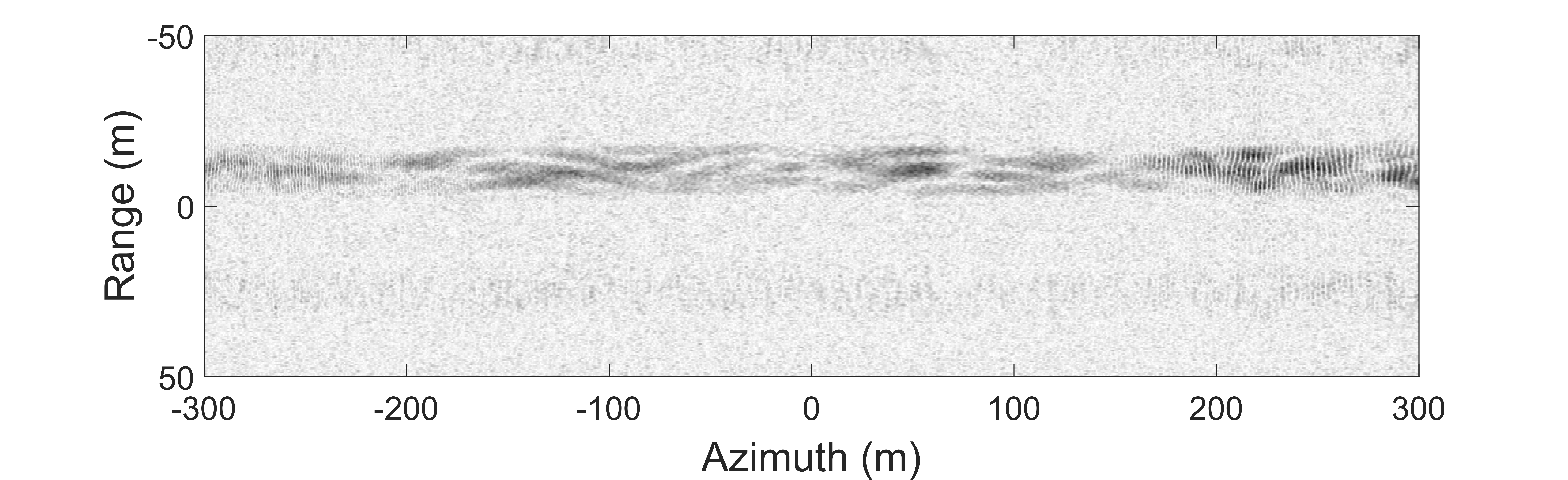}%
\label{fig4_b}}
\vfil
\subfloat[Result by the compensation based method]{\includegraphics[width=2.5in]{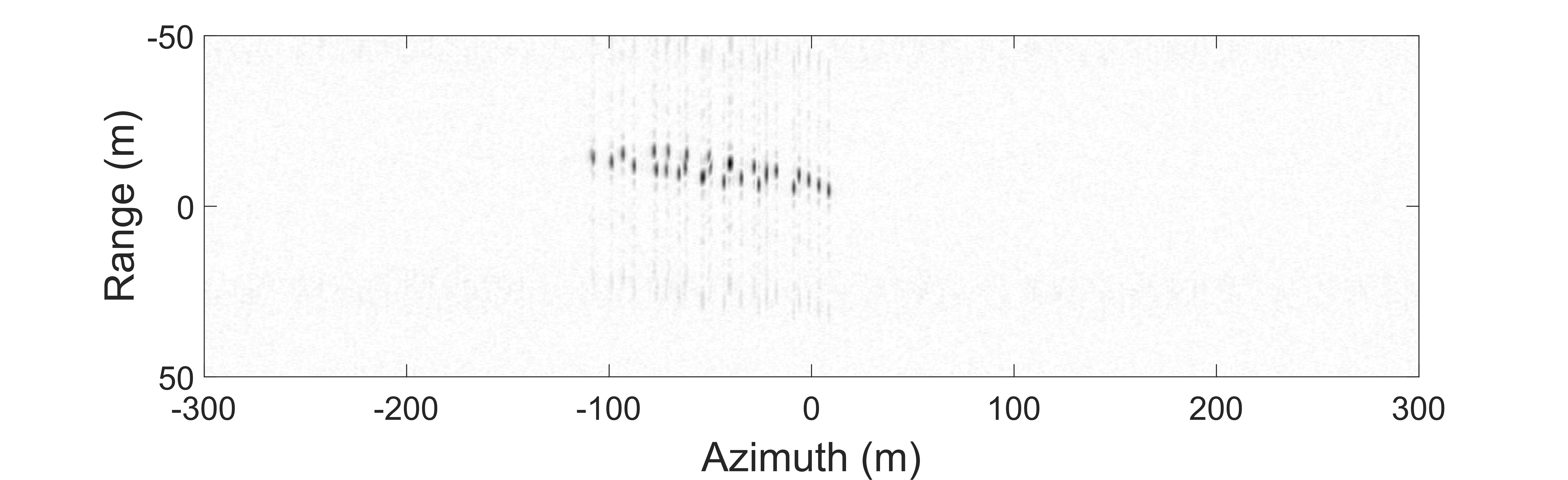}%
\label{fig4_b}}
\caption{Simulation result of ship target. (a) Original dot matrix of ship target. (b) Result by Range-Doppler algorithm. (c) Result by the proposed method.}
\label{f5}
\end{figure}

This example images a dot matrix of ship target with nonstationary motion. The simulation result demonstrates the proposed JRC receiver’s ability to distinguish and focus separated, nonstationary targets influenced by ocean waves. The geometry and system configurations are the same as the previous simulation. The original dot matrix and imaging result are shown in Fig. \ref{f5}. 

\section{Conclusion}
In this paper, the feasibility of applying a JRC waveform in GEO SA-Bi SAR ship target imaging task is investigated. A JRC receiver which located on the airborne SAR platform is built to perform communication and SAR imaging simultaneously. A previously developed SAR nonstationary target imaging method is extended and employed to established this JRC strategy. The simulation results demonstrate the correctness and effectiveness of the proposed JRC strategy. In future work, the the applicability of artificial neural network on JRC design will be investigated.

\end{document}